\documentclass[preprint,5p,times]{elsarticle}
\usepackage{epsfig}
\usepackage{amssymb}
\usepackage{amsthm}

\newcommand{\pol}[1]{\mathaccent"017E{#1}}

\journal{Physics Letters B}

\begin{document}

\begin{frontmatter}

\title{Measurement of the analysing power in proton-proton elastic scattering at small
angles}

\author[hepi,ikp]{Z.~Bagdasarian\corref{cor1}}
\ead{z.bagdasarian@fz-juelich.de} 
\cortext[cor1]{Corresponding author.}

\author[hepi,ikp]{D.~Chiladze}
\author[jinr1,ikp]{S.~Dymov}
\author[ikp]{A.~Kacharava}
\author[hepi,jinr1]{G.~Macharashvili}

\author[pnpi]{S.~Barsov}
\author[ikp]{R.~Gebel}
\author[cas,ikp]{B.~Gou}
\author[ikp]{M.~Hartmann}
\author[ikp]{I.~Keshelashvili}
\author[munster]{A.~Khoukaz}
\author[krakow]{P.~Kulessa}
\author[jinr1]{A.~Kulikov}
\author[ikp]{A.~Lehrach}
\author[hepi]{N.~Lomidze}
\author[ikp]{B.~Lorentz}
\author[ikp]{R.~Maier}
\author[hepi,ikp]{D.~Mchedlishvili}
\author[jinr1,ikp]{S.~Merzliakov}
\author[ikp,pnpi]{S.~Mikirtychyants}
\author[hepi]{M.~Nioradze}
\author[ikp]{H.~Ohm}
\author[munster]{M.~Papenbrock}
\author[ikp]{D.~Prasuhn}
\author[ikp]{F.~Rathmann}
\author[ikp]{V.~Serdyuk}
\author[jinr1]{V.~Shmakova}
\author[ikp]{R.~Stassen}
\author[ikp]{H.~Stockhorst}
\author[wash]{I.I.~Strakovsky}
\author[ikp]{H.~Str\"oher}
\author[hepi]{M.~Tabidze}
\author[munster]{A.~T\"{a}schner}
\author[ikpros,msu]{S.~Trusov}
\author[jinr1]{D.~Tsirkov}
\author[jinr1]{Yu.~Uzikov}
\author[pnpi,bonn]{Yu.~Valdau}
\author[ucl]{C.~Wilkin}
\author[wash]{R.L.~Workman}

\address[hepi]{High Energy Physics Institute, Tbilisi State University, GE-0186 Tbilisi, Georgia}
\address[ikp]{Institut f\"ur Kernphysik, Forschungszentrum J\"ulich, D-52425
 J\"ulich, Germany}
\address[jinr1]{Laboratory of Nuclear Problems, JINR, RU-141980 Dubna, Russia}
\address[pnpi]{High Energy Physics Department, Petersburg Nuclear Physics
Institute, RU-188350 Gatchina, Russia}
\address[cas]{Institute of Modern Physics, Chinese Academy of Sciences, Lanzhou 730000, China}
\address[munster]{Institut f\"ur Kernphysik, Universit\"at M\"unster, D-48149
M\"unster, Germany}
\address[krakow]{H.~Niewodnicza\'{n}ski Institute of Nuclear Physics PAN, PL-31342
Krak\'{o}w, Poland}
\address[wash]{Data Analysis Center at the Institute for Nuclear Studies,
Department of Physics,\\
The George Washington University, Washington,D.C. 20052, USA}
\address[ikpros]{Institut f\"ur Kern- und Hadronenphysik,
Forschungszentrum Rossendorf, D-01314 Dresden, Germany}
\address[msu]{Department of Physics, M.~V.~Lomonosov Moscow State University,
RU-119991 Moscow, Russia}
\address[bonn]{Helmholtz-Institut f\"ur Strahlen- und Kernphysik, Universit\"at
  Bonn, D-53115 Bonn, Germany}
\address[ucl]{Physics and Astronomy Department, UCL, London WC1E 6BT, UK}

\begin{abstract}
%% Text of abstract
The proton analysing power in $\pol{p}p$ elastic scattering has
been measured at small angles at COSY-ANKE at 796~MeV and five
other beam energies between 1.6 and 2.4~GeV using a polarised
proton beam. The asymmetries obtained by detecting the fast
proton in the ANKE forward detector or the slow recoil proton
in a silicon tracking telescope are completely consistent.
Although the analysing power results agree well with the many
published data at 796~MeV, and also with the most recent
partial wave solution at this energy, the ANKE data at the
higher energies lie well above the predictions of this solution
at small angles. An updated phase shift analysis that uses the
ANKE results together with the World data leads to a much
better description of these new measurements.
\end{abstract}

\begin{keyword}

Proton-proton elastic scattering \sep analysing power \sep phase shift analysis

\PACS 13.75.Cs 	 %Nucleon-nucleon interactions (including antinucleons, deuterons, etc.)
\sep 24.70.+s    %Polarisation phenomena in reactions
\sep 25.40.Cm 	 %Elastic proton scattering

\end{keyword}

\end{frontmatter}

%% \linenumbers

The measurements of proton-proton elastic scattering undertaken
by the COSY-EDDA collaboration have had a major impact on the
partial wave analysis of this reaction above
1~GeV~\cite{ARN2000}. The data on the differential cross
section~\cite{ALB1997} were taken in a continuous ramp from
0.24 to 2.58~GeV and analogous results were produced for the
proton analysing power between 0.44 and
2.49~GeV~\cite{ALT2005}. In addition, $pp$ spin correlations
were studied between 0.48 and 2.49~GeV~\cite{BAU2003}. However,
due to the design of the EDDA detector, the experiments only
extended over the central region of centre-of-mass (c.m.)
angles, $30^{\circ} \lesssim \theta_{cm} \lesssim 150^{\circ}$,
and there are very few other analysing power measurements
available below $30^{\circ}$ for beam energies above
1~GeV~\cite{ARN2000}. The lack of data has left major
ambiguities in the phase shift analysis. In complete contrast
to COSY-EDDA, the COSY-ANKE facility was designed for the
investigation of the small angle region and is thus well suited
to cover this significant gap in the database.

\begin{figure}[h]
\includegraphics[width=1.0\linewidth]{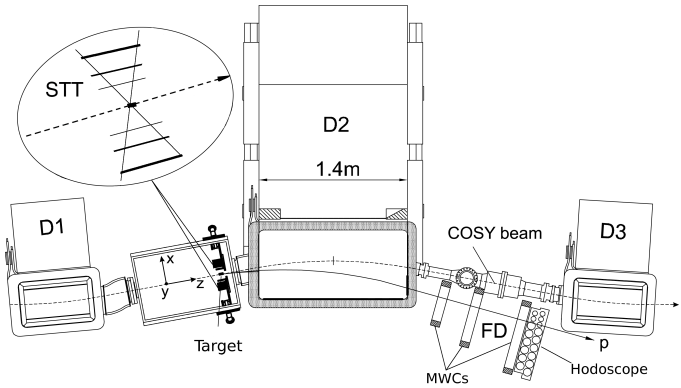}
\caption{The ANKE spectrometer setup (top view), showing the positions of
the hydrogen cluster-jet target, the silicon tracking
telescopes (STT), and the forward detector (FD).
}
\label{ankesetup}
\end{figure}

The present experiment was carried out using the ANKE magnetic
spectrometer~\cite{BAR2001} positioned inside the storage ring of the COoler
SYnchrotron (COSY)~\cite{MAI1997} of the Forschungs\-zentrum J\"ulich.
Although the facility sketched in Fig.~\ref{ankesetup} is equipped with other
elements, the only detectors used in this experiment were the forward
detector (FD) and the silicon tracking telescopes (STT)~\cite{SCH2003}.

The fast protons from elastic $pp$ scattering were measured in
the forward detector which, for $pp$ elastic scattering,
covered $10^{\circ}-30^{\circ}$ in c.m.\ polar angles and $\pm
30^{\circ}$ in azimuth. The FD comprises a set of multiwire
proportional and drift chambers (MWCs) and a two-plane
scintillation hodoscope. The counters were used to measure the
energy losses required for particle
identification~\cite{DYM2004}.

The two STT were placed symmetrically inside the vacuum
chamber, to the left and right of the beam near the unpolarised
hydrogen cluster-jet target~\cite{KHO1999}. Each telescope
consists of three sensitive silicon layers of 70~$\mu$m,
300~$\mu$m, and 5~mm thickness and covers the laboratory polar
angles $75^{\circ} < \theta_{\rm lab} < 140^{\circ}$. In order
to pass through the three layers, the protons must have kinetic
energies of at least 2.5~MeV, 6~MeV, and 30~MeV, respectively.
For stopping protons with energies below 30~MeV the particle
identification is unambiguous.  In this case greater precision
in the angle of the recoiling proton is achieved by deducing it
from the energy measured in the telescope rather than from a
direct angular measurement. However, by studying the energy
deposited principally in the third layer, it is also possible
to deduce the energy of punch-through protons up to 90~MeV,
thus expanding considerably the angular coverage of the
telescope. For this purpose the kinetic energy of these fast
protons was defined through a comparison of the angles and
energy deposits with simulated data using a neural network
approach~\cite{ANK2014}.

Two independent triggers were used in the determination of the analysing
powers. The FD trigger required a coincidence between the two planes in the
hodoscope while the STT trigger requested a minimum energy deposit in the
second layer of either telescope. Due to the overlap in the angular
acceptance, some events were registered with both triggers. However, since
the two data sets were then analysed independently, this did not bias the
either set of results.

The ANKE experiment used a vertically polarised beam incident
on an unpolarised target so that the preparation and the
measurement of the beam polarisation are critical. The H$^-$
ions from the polarised ion source were accelerated to 45~MeV
in the cyclotron JULIC before being stripped of their electrons
and injected into COSY~\cite{EVE1993}. Two modes, with spin up
$(\uparrow)$ and down $(\downarrow)$, were supplied by the
source and the polarisations of the injected beam were
optimised using a low energy polarimeter (LEP) in the injection
beam line to COSY~\cite{CHI2006}. The LEP measurements showed
that the magnitudes of the polarisations were typically about
93\% and the difference between the values of the two modes was
smaller than the statistical uncertainty of 1\%.

In a strong-focusing synchrotron, such as COSY, resonances can
lead to losses of polarisation of a proton beam during
acceleration. In order to compensate for these effects,
adiabatic spin-flip was used to overcome the imperfection
resonances and tune-jumping to deal with the intrinsic
ones~\cite{LEH2003}. The polarisations were measured using the
EDDA detector as a polarimeter. This detector, originally
equipped with a polarised hydrogen target, had been used to
measure the $pp$ analysing power over almost the whole COSY
energy range ~\cite{ALT2005}. By studying further the
scattering of polarised protons on C and CH$_2$ targets, it was
possible to deduce the quasi-free analysing power of the
carbon, where the necessary calibration standard was provided
by the EDDA $p\pol{p}$ data~\cite{WEI2000}.

The stripped-down version of the EDDA detector used as a
polarimeter at COSY was calibrated during the EDDA data-taking
periods against the full detector setup.  The 7~$\mu$m diameter
carbon fibre target is moved into the beam from below. The
polarimeter consists of 29 pairs of half-rings placed to the
left and right of the beam. It is therefore possible to compare
the rates in the left and right half-rings for each range in
polar angle $\theta_{\rm lab}$ while averaging over the
azimuthal angle $\phi$ in every half-ring. In order to assure
fast polarimetry, the coincidences are recorded by scalers. The
asymmetry is determined individually for each pair of
half-rings and the weighted average evaluated. The systematic
uncertainty of the measurements was estimated to be 3\% at each
energy~\cite{WEI2000}.

The experiment at ANKE was carried out at six energies,
$T_p=796$, 1600, 1800, 1965, 2157, and 2368~MeV. Cycles of
180~s or 300~s duration were used for each spin mode, with the
last 20~s of each cycle being reserved for the measurement of
the beam polarisation with the EDDA detector\footnote{The EDDA
target effectively consumes all the beam so that it could not
be used before an ANKE measurement in a cycle.}. Consistent
results were achieved with EDDA after the short and long cycles
which, as expected, implies that beam polarisation is not lost
over a COSY cycle. However, due to the non-zero dispersion
combined with the energy loss of the beam caused by its passage
through the target, the settings at the three lowest energies
gradually degrade slightly. This effect was taken into account
in the analysis.
\begin{table}[h!]
\caption{The values of the mean polarisations $p$ determined
with the EDDA polarimeter averaged over all the data at the
beam energy $T_p$ where the $pp$ analysing power was measured
in ANKE. The changes in sign in $p$ are due to the spin flips
induced when passing through the imperfection resonances.
Though the shown statistical errors are small, there are 3\%
systematic uncertainties~\cite{WEI2000}. The normalisation
factors $N$ are those obtained in a partial wave
fit~\cite{ARN2000} to the current STT data, as discussed in the
text. \label{polar} \vspace{3mm} }%

\centering
\scriptsize{
\begin{tabular}{|c|c|c|c|c|c|c|}
\hline
$T_p$ (MeV)&796&1600&1800&1965&2157&2368\\
\hline
$p$       & \,0.554\,&\,0.504\,&\,$-0.508$\,&\,$-0.429$\,&\,$-0.501$\,&\,$0.435$\,\\
          & $\pm0.008$ & $\pm0.003$ & $\pm0.011$ & $\pm0.008$ & $\pm0.010$ & $\pm 0.015$\\
\hline
$N$&1.00&1.00&0.99&1.09&1.01&0.93\\
\hline
\end{tabular}
}
\end{table}

The weighted averages over time and polar angle of the beam
polarisations determined using the EDDA polarimeter at the six
energies are given in Table~\ref{polar}. The values correspond
to half the difference between spin up and down data and the
changes in sign reflect the number of spin flips required to
pass through the imperfection resonances. The variation of the
beam polarisation cycle by cycle was checked with the asymmetry
of the counts in STT and found to be around 0.04 (RMS). It should be noted that, even for the lowest energy of 796 MeV, two 
intrinsic and two imperfection resonances have to be crossed during the 
acceleration and this results in polarisations of less than 60\% for all 
the energies investigated. At each of these six energies the beams were 
prepared independently and, for this reason, the magnitude of the 
polarisation may not decrease monotonically as more resonances are 
crossed.

\begin{figure}[h]
\includegraphics[width=1.00\linewidth]{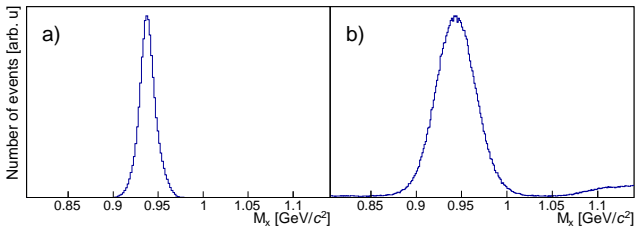}
\caption{Missing-mass $M_X(pp\to pX)$ spectra obtained for a
beam energy of 1.6~GeV showing the clear proton peak when
detecting one proton in (a) the STT and (b) the FD.}
\label{missing_mass}
\end{figure}

In the ANKE experiment a proton is measured in either the STT
or FD and elastic $pp$ scattering events identified through the
evaluation of the missing mass in the reaction. As can be seen
from typical examples of both cases shown in
Fig.~\ref{missing_mass} at a beam energy of 1.6~GeV, there is
very little ambiguity in the isolation of the proton peak. The
greater suppression of events associated with pion production
in the STT is due to the minimum longitudinal momentum of the
recoil proton and the restricted angular acceptance of this
detector.

The left/right symmetry of the STT system reduces some of the
systematic uncertainties. The so-called cross-ratio
method~\cite{OHL1973} allows one to eliminate first-order
systematic errors that arise from misalignments between the two
STT and for this reason the beam polarisation was reversed in
each successive cycle. Let $L_\uparrow (L_\downarrow)$ be the
numbers of counts in the left telescope with spin up (down) and
$R_\uparrow (R_\downarrow)$ the analogous quantities for the
right telescope. In terms of the geometric means,
$L=\sqrt{L_\uparrow R_\downarrow}$ and $R=\sqrt{R_\uparrow
L_\downarrow}$, the scattering asymmetry is related to the
analysing power $A_y(\theta)$ for each value of the scattering
angle $\theta$ through
\begin{equation}
\varepsilon(\theta)=\frac{L(\theta)-R(\theta)}{L(\theta)+R(\theta)}
=A_y(\theta)\,p\,\langle \cos\phi \rangle,
\end{equation}
where $p\,\langle \cos\phi \rangle$ is the effective beam polarisation,
taking into account the acceptance of the STT in the azimuthal angle $\phi$.
In our geometry $\langle\cos\phi\rangle\approx 0.966$.

Other systematic errors, such as those arising from differences
in the magnitudes of the up and down polarisations, also cancel
in first order. The overall systematic uncertainty in $A_y$
arising from asymmetry measurement with STT does not exceed
0.3\%. Another factor that could affect the asymmetry measured
with such a two-arm detector is any instability in the ratio of
the efficiencies of the left and right telescopes. The
instability correction, which was studied at all energies, does
not exceed the $|c|=1.3\%$ that was found at 1.8~GeV. The
relevant corrections of the analysing power
$c(\theta)A_{y}(\theta)$ were added for each angular
bin~\cite{ANK2014}.

The absence of the left-right symmetry in the forward detector
does not permit the use of the cross-ratio method, and the
analysing power can only be defined from the asymmetry of the
count-rates for the two states of the beam polarisation. The
number of events for each orientation  of the polarisation was
weighted with the relative luminosity factors, which were fixed
by comparing the rates of charged particle production in
angular regions where the beam polarisation could play no
part~\cite{YAS2005}. Since the calibration events were selected
with the same trigger as that used for $pp$ elastic scattering,
this procedure automatically takes into account any dead-time
difference between the spin-up and spin-down data. The
calibration data, which corresponded generally to inelastic
events involving pion production, were selected by putting cuts
either on small polar angles $\theta$ or on the azimuthal angle
$\phi$ near $\pm 90^{\circ}$. Consistent values for the
relative luminosities were achieved when varying these cuts and
it is estimated that the systematic uncertainty of $A_y$ due to
the relative luminosity normalisation never exceeds 0.3\%. This
approach could be checked by comparing the FD and STT results
in the angular overlap regions.

The efficiency for registering events in the forward detector
induced by spin-up or spin-down protons was studied by using
events where both the fast and recoil protons were measured in
the FD and STT, respectively. The differences of the
efficiencies of less than $10^{-3}$ could be neglected compared
to the statistical uncertainties. Potentially more serious for
the FD analysis is the assumption that the magnitudes of the
two polarisation modes were identical, viz.\
$|p_{\uparrow}|=|p_{\downarrow}|$. Whereas deviations from the
mean are very small at injection, and are known to be less than
5\% after acceleration, these could induce fractional errors in
$A_y$ of up to 2.5\%. It should, however, be remarked that in
the overlap regions of the STT and FD data any disagreements
between the determinations of the asymmetries in the two
systems are on the 1\% level and this puts a much tighter
constraint on possible $|p_{\uparrow}|,\ |p_{\downarrow}|$
differences.

\begin{figure*}[htb]
\includegraphics[width=0.99\linewidth,angle=0]{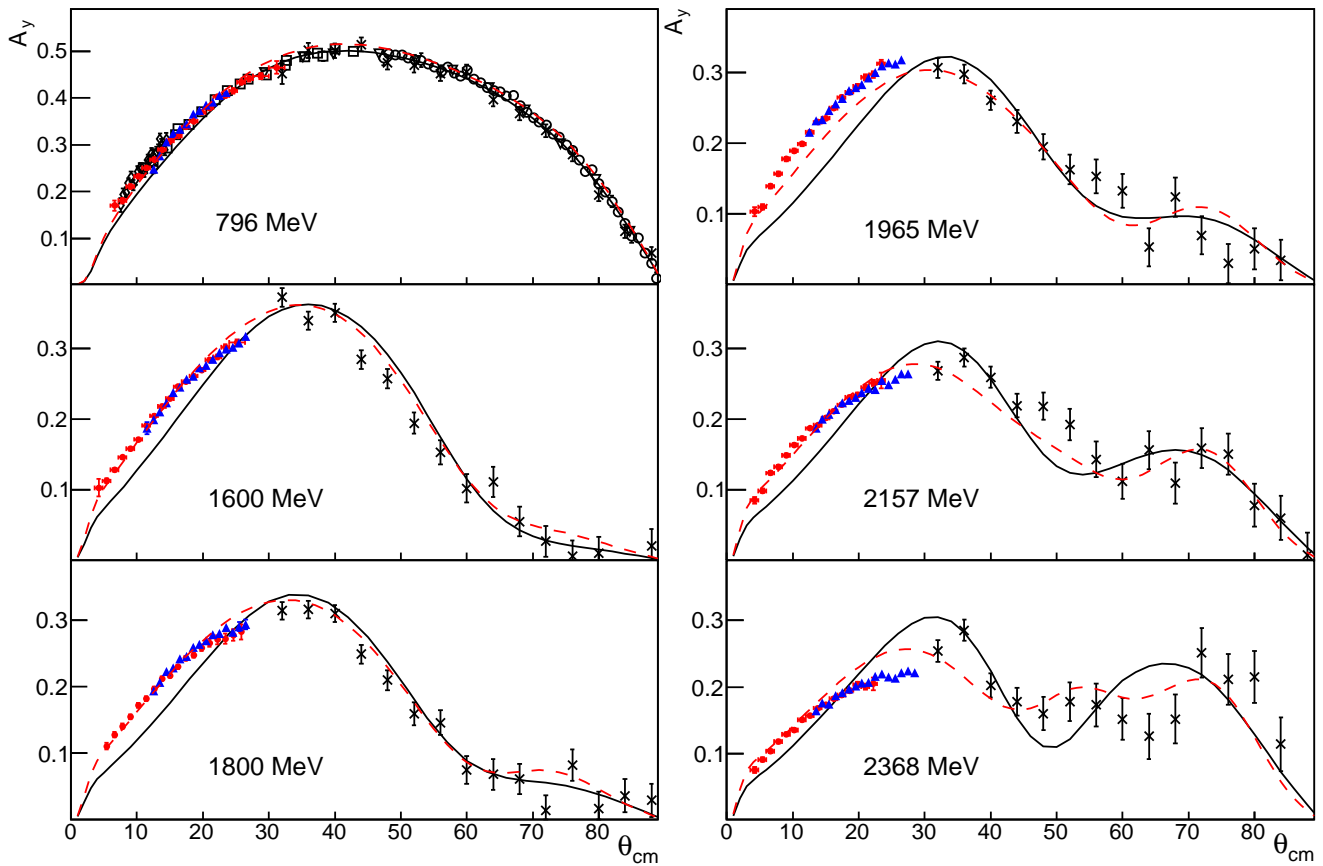}
\caption{(Color online) Comparison of the ANKE measurements of the proton
analysing power in $pp$ elastic scattering using the STT (red filled circles)
and FD (blue filled triangles) systems with the curves corresponding to the
SAID SP07 (solid black line) and the revised fit (dashed red)
solutions~\cite{ARN2000}. Only statistical errors are shown so that the
systematic uncertainties arising, for example, from the calibration of the
EDDA polarimeter have not been included. Also shown are selected results from
EDDA (black crosses)~\cite{ALT2005} at the energies different by no more than
7 MeV  and, at 796~MeV, LAMPF ~\cite{MCN1981,IRO1982,BEV1978}, and SATURNE
~\cite{ALL1998} (black open symbols). It should be noted that the EDDA data
were taken over a continuous ramp~\cite{ALT2005} and, if continuity in energy
were imposed, many of their statistical fluctuations would be diminished.}
\label{money_plot}
\end{figure*}

There is also a systematic uncertainty in the determination of
the scattering angle, and this could affect both the STT and FD
data. The simultaneous measurement of the deuteron and pion
from the $pp\to d\pi^+$ reaction in the forward detector showed
that the systematic deviations in the laboratory angles from
those expected for these kinematics did not exceed
$0.07^{\circ}$. If this is valid also for $pp$ elastic
scattering it would suggest that the c.m.\ scattering angles
were defined with a precision of better than $0.15^{\circ}$.

In cases where one of the protons from an elastic scattering event is
detected in the FD and the other in the STT it is possible to compare
directly the scattering angle measured in the two systems. In general
$\theta_{cm}({\rm STT}) > \theta_{cm}({\rm FD})$, with the difference being
typically $\approx 0.3^{\circ}$. It is not possible to judge which detector
is responsible for this difference which is, however, small compared with the
bin widths of $1.0^{\circ}$ (FD) and $1.2^{\circ}$ (STT).

The dominant systematic error is that arising from the determination of the
beam polarisation in the EDDA polarimeter, which was estimated to be
3\%~\cite{WEI2000}. For the FD data there is, in addition, a possible
contribution associated with the assumption of equal up and down
polarisations so that in this case we would cautiously assume a 5\%
systematic uncertainty. To these figures must be added the statistical
uncertainty in the determinations of the beam polarisations at the six
energies shown in Table~\ref{polar}.

The results of all the ANKE measurements of $A_y$ for $pp$
elastic scattering are shown for the six energies in
Fig.~\ref{money_plot}. The agreement between the STT and FD
data, which involved completely independent measurements of the
final state, is remarkably good. The individual deviations
generally lie within the statistical error bars and the average
over the angular overlap regions is $A_{y}({\rm FD})/A_{y}({\rm
STT})=1.00\pm 0.01$. At beam energies close to 796~MeV there
are many measurements of the $pp$ analysing power and, in
general, they are in good mutual agreement, as they are with
the new ANKE data. This reinforces the confidence in the use of
the EDDA polarimeter. At 1.6~GeV and above there are far fewer
experimental measurements and, for clarity, we only show the
EDDA data at neighbouring energies though, at the highest
energy, the statistical fluctuations are
significant~\cite{ALT2005}.

The SAID SP07 solution~\cite{ARN2000}, shown by the solid black
line in Fig.~\ref{money_plot}, describes the bulk of the
$\approx 796$~MeV data very well indeed. However, at higher
energies the ANKE data deviate significantly from the
predictions of the SP07 solution. Moreover, the shapes of the
ANKE data seem very different from these predictions, rising
much more steeply at small angles. Therefore, these
discrepancies cannot be due to a simple miscalibration of the
EDDA polarimeter, for example, which would change the overall
magnitude of $A_y(\theta)$ but not its angular dependence.

The ANKE analysing power data have been added to the World data
set and searches made for an updated partial wave
solution~\cite{ARN2000}. To allow for possible systematic
effects, the SAID fitting procedure introduces a scale factor
$N$ into any data set and determines its value, as well as the
$pp$ phases and inelasticities, by minimising an overall
$\chi^2$ for the complete data set. When this is done, the
average value of $\chi^2$ per degree of freedom found for the
ANKE STT data is 1.6 and slightly larger for the FD results.
The new fits, which lead to the red dashed curves in
Fig.~\ref{money_plot}, correspond to relatively modest changes to the parameters for several
of the lower partial waves, with the biggest change being in $^{3}F_2$. The values of the
normalisation factors $N$ reported in Table~\ref{polar} have an
average of $\langle N\rangle = 1.00\pm 0.02$ for the STT data.
These factors, which would effectively multiply the beam
polarisations, have not been applied in Fig.~\ref{money_plot}.
The deviations of the  individual values of $N$ from unity
might seem to be greater at the higher energies. They are
somewhat larger than what one would expect on the basis of the
quoted uncertainties in the EDDA polarimeter, being around 5\%
rather than the 3\% estimate~\cite{WEI2000}. It should be
stressed that the introduction of the scale factor $N$ does not
change the shape of a distribution and, even in cases where a
value close to one is found, this does not mean that the fit
reproduces perfectly the data. A clear example of this is to be
found in the larger angle data at 1.6~GeV shown in
Fig.~\ref{money_plot}.

In summary we have measured the analysing power in $pp$ elastic
scattering at 796~MeV and at five energies from 1.6~GeV up to
2.4~GeV using both the silicon tracking telescopes and the ANKE
forward detector. The consistency between these two independent
measurements of the final protons is striking so that the only
major systematic uncertainty is the few percent associated with
the calibration of the EDDA polarimeter. Though the overall
uncertainties are slightly larger for the FD data, these
results are important because they extend the coverage to
slightly larger scattering angles.

In the small angle range accessible to ANKE, the new data are
consistent with older measurements around 796~MeV and also with
the SP07 SAID predictions at this energy~\cite{ARN2000}. At
higher energies the ANKE results lie significantly above the
SP07 solution near the forward direction and also display a
different angular dependence. By adjusting some of the phases
and inelasticities in the low partial waves of this solution it
has been possible to obtain a much better description of the
ANKE $A_y$ data with reasonable values of
$\chi^2/\textit{NDF}$. However, this is at the expense of
introducing renormalising factors that deviate from unity by
more than expected on the basis of the estimated systematic
uncertainties arising from the use of the EDDA polarimeter. The
situation may be changed somewhat when the corresponding
unpolarised differential cross sections~\cite{CHI2011} become
available since these data, which cover rather unexplored
regions, might modify the parameters of the ``best'' partial
wave solution. These ANKE data are still being processed.

We acknowledge valuable discussions with Frank Hinterberger. We
also are grateful to other members of the ANKE Collaboration
for their help with this experiment and to the COSY crew for
providing such good working conditions. This work has been
supported by the Forschungszentrum J\"ulich (COSY-FEE), by the
U.S.\ Department of Energy, Office of Science, Office of
Nuclear Physics, under Award Number DE.FG02.99ER41110, and by
the Shota Rustaveli Science Foundation Grant.

\end{document}